\documentclass[12pt,epsf]{article}
\usepackage{verbatim}

\usepackage{dsfont}
\usepackage{sidecap}
\sidecaptionvpos{figure}{t}
\usepackage{subfigure}

\usepackage[english]{babel}

\usepackage{amssymb,amsmath}
\usepackage{graphicx, xcolor, varwidth}
\usepackage{setspace}
\usepackage[permil]{overpic}
\usepackage{cite}

\usepackage{feynmp}
\colorlet{darkblue}{blue!70!black}
\colorlet{darkgreen}{green!70!black}

\usepackage[colorlinks=true,urlcolor=darkblue,linktocpage=true,linkcolor=darkblue,citecolor=darkblue]{hyperref}

\numberwithin{equation}{section}

\newcommand{\be}{\begin{equation}}
\newcommand{\ee}{\end{equation}}
\newcommand{\bea}{\begin{eqnarray}}
\newcommand{\eea}{\end{eqnarray}}
\newcommand{\bear}{\begin{eqnarray}}
\newcommand{\eear}{\end{eqnarray}}  
\newcommand{\beas}{\begin{eqnarray*}}
\newcommand{\p}{\partial}
\newcommand{\eeas}{\end{eqnarray*}}
\newcommand{\ba}{\begin{array}}
\newcommand{\ea}{\end{array}}

\newcommand{\del}{\nabla}

\newcommand{\pd}[2][1]{\ifnum#1=1 \frac{\partial}{\partial {#2}} \else
  \frac{\partial^#1}{\partial {#2}^{#1}}\fi}
\newcommand{\dpd}[2][1]{\ifnum#1=1 \dfrac{\partial}{\partial {#2}} \else
  \frac{\partial^#1}{\partial {#2}^{#1}}\fi}
\newcommand{\td}[2][1]{\ifnum#1=1 \frac{d}{d{#2}} \else
  \frac{d^#1}{d{#2}^{#1}}\fi}

\newcommand{\e}{\varepsilon}

\newcommand{\nbox}{{\,\lower0.9pt\vbox{\hrule \hbox{\vrule height 0.2 cm \hskip 0.19 cm \vrule height 0.2 cm}\hrule}\,}}

\newcommand{\ie}{{\it i.e.,}\ }

\newcommand{\bh}{\bar{h}}

\newcommand{\half}{\tfrac{1}{2}}

\newcommand{\bz}{\bar{z}}

\newcommand{\tn}{\tilde{n}}



\renewcommand{\e}{\epsilon}
\newcommand{\sdot}{\!\cdot\!}

\newdimen\tableauside\tableauside=1.0ex
\newdimen\tableaurule\tableaurule=0.4pt
\newdimen\tableaustep
\def\phantomhrule#1{\hbox{\vbox to0pt{\hrule height\tableaurule width#1\vss}}}
\def\phantomvrule#1{\vbox{\hbox to0pt{\vrule width\tableaurule height#1\hss}}}
\def\sqr{\vbox{%
  \phantomhrule\tableaustep
  \hbox{\phantomvrule\tableaustep\kern\tableaustep\phantomvrule\tableaustep}%
  \hbox{\vbox{\phantomhrule\tableauside}\kern-\tableaurule}}}
\def\squares#1{\hbox{\count0=#1\noindent\loop\sqr
  \advance\count0 by-1 \ifnum\count0>0\repeat}}
\def\tableau#1{\vcenter{\offinterlineskip
  \tableaustep=\tableauside\advance\tableaustep by-\tableaurule
  \kern\normallineskip\hbox
    {\kern\normallineskip\vbox
      {\gettableau#1 0 }%
     \kern\normallineskip\kern\tableaurule}%
  \kern\normallineskip\kern\tableaurule}}
\def\gettableau#1 {\ifnum#1=0\let\next=\null\else
  \squares{#1}\let\next=\gettableau\fi\next}

\renewcommand{\e}{\varepsilon}

\begin{document}
\begin{spacing}{1.3}
\begin{titlepage}

\begin{center}
{\Large \bf A New Spin on Causality Constraints}

\vspace*{12mm}

Thomas Hartman, Sachin Jain, and Sandipan Kundu
\vspace*{6mm}

\textit{Department of Physics, Cornell University, Ithaca, New York\\}

\vspace{6mm}

{\tt hartman@cornell.edu, sj339@cornell.edu, kundu@cornell.edu}

\vspace*{15mm}
\end{center}
\begin{abstract}

Causality in a shockwave state is related to the analytic properties of a four-point correlation function. Extending recent results for scalar probes, we show that this constrains the couplings of the stress tensor to light spinning operators in conformal field theory, and interpret these constraints in terms of the interaction with null energy. For spin-1 and spin-2 conserved currents in four dimensions, the resulting inequalities are a subset of the Hofman-Maldacena conditions for positive energy deposition. It is well known that energy conditions in holographic theories are related to causality on the gravity side; our results make a connection on the CFT side, and extend it to non-holographic theories.

\end{abstract}

\end{titlepage}
\end{spacing}

\vskip 1cm

\setcounter{tocdepth}{2}
\tableofcontents

\begin{spacing}{1.3}
\newpage 

\section{Introduction}

We recently showed that causality imposes inequalities on coupling constants in strongly interacting conformal field theory in $d>2$ dimensions \cite{paper1}. The constraints for scalar correlators, derived by conformal bootstrap methods, were nontrivial only in special cases, but we argued that similar logic would impose nontrivial constraints on the couplings of spinning operators.  In this paper we confirm this, deriving a set of causality constraints on the couplings of spinning operators that hold in any unitary CFT in $d>2$ dimensions (up to a caveat below). Like the causality constraints of \cite{Adams:2006sv}, or the results of the computational conformal bootstrap \cite{Rattazzi:2008pe, ElShowk:2012ht}, these are infrared constraints imposed by ultraviolet consistency.

\subsection{Energy condition from causality}
We study the 4-point function $\langle \psi OO \psi\rangle$, where $\psi$ is a scalar and $O$ is a symmetric traceless tensor operator of spin $\ell$. The spinning operator $O$ is viewed as a `probe' in the background produced by $\psi$. Causality requires
\be\label{introc}
\langle \psi(x_1)\,  \big[O_{\mu_1 \dots \mu_\ell}(x_2)\ ,\   O_{\nu_1 \dots \nu_\ell}(x_3)\big]\,  \psi(x_4)\rangle = 0
\ee
for spacelike separated probes, $(x_2-x_3)^2 > 0$. We show that vanishing of this commutator imposes a constraint on the coupling to null energy, in a limit with lightlike operator insertions:
\be\label{nulle}
\frac{ \langle \varepsilon\sdot O(0) \varepsilon^* \sdot O(y^+,y^-) T_{--}(\infty \hat{y})\rangle }{\langle \varepsilon\sdot O(0) \varepsilon^* \sdot O(y^+,y^-) \rangle} < 0
\quad \mbox{as} \quad y^+ \to 0 \ ,
\ee
with $y^\pm = y \pm t > 0$. 
This particular 3-point coupling appears because it determines the leading behavior of the conformal block for stress tensor exchange, $OO \to T \to \psi\psi$, in the lightcone limit. Note the peculiar sign of the constraint --- our sign conventions are such that the usual null energy condition is $\langle T_{--} \rangle >0$. Of course this is not a contradiction, since \eqref{nulle} is not an expectation value.

Unlike the null energy condition, which is violated quantum mechanically, the energy condition \eqref{nulle} is derived from reflection positivity of 4-point functions in the Euclidean theory and must hold in any unitary CFT in $d>2$ obeying the usual axioms, with one notable caveat: We assume that no scalar fields with scaling dimension in the range $\frac{d}{2}-1 < \Delta \leq d-2$ appear in the $OO$ OPE. This ensures that stress tensor exchange dominates the 4-point function \eqref{introc} for null-separated probes.\footnote{In several interesting CFTs, including the Wilson-Fisher and Klebanov-Witten fixed points, there is a scalar in this range but symmetries prevent it from appearing in this OPE.  See \cite{Bousso:2014uxa} for a discussion in the context of entanglement entropy (which has an enhanced divergence if such operators appear in the $TT$ OPE). }

If \eqref{nulle} is violated, then the 4-point function $\langle \psi OO\psi\rangle$ has a singularity before the Minkowski lightcone, and therefore a non-vanishing commutator at spacelike separation. In other words, the shockwave induces a time advance.

\subsection{Currents and stress tensors in $d=4$}

The complete 3-point function $\langle OOT\rangle$ is fixed by conformal symmetry up to numerical couplings, so \eqref{nulle} is a set of inequalities for these coupling constants. Different polarizations lead to multiple, independent constraints. We will work out these constraints in detail for the case of spin-1 and spin-2 conserved probes in four spacetime dimensions.

A convenient basis for 3-point tensor structures of conserved currents is provided by free fields of various spins. (We assume parity.) The current 3-point functions in any CFT can be expanded in this basis,
\be
\langle JJT \rangle = n_s \langle JJT\rangle_{scalar} + n_f \langle JJT\rangle_{fermion}
\ee
and
\be
\langle TTT\rangle = \tn_s\langle TTT\rangle_{scalar} + \tn_f \langle TTT\rangle_{fermion} + \tn_v\langle TTT\rangle_{vector} \ ,
\ee
where $\langle OO T \rangle_i$ is proportional to the correlator in a theory of free scalars, fermions, or vectors, and $n_s, n_f, \tn_s, \tn_f, \tn_v$ are the coupling constants (see appendices for conventions). 

The conformal Ward identity relates one combination of couplings to the two-point function, which must be positive in a unitary theory. With our normalizations, these combinations are 
\be\label{wardc}
c_J \propto 4\tn_f  + \frac{\tn_s}{d-2}  >  0  \ , \qquad  c_T \propto \tn_f + \frac{1}{2(d-1)}\tn_s + \frac{16(d-3)}{d(d-2)}\tn_v  >   0  \ .
\ee
Hofman and Maldacena \cite{Hofman:2008ar,Hofman:2009ug} (and others in $d \neq 4$ \cite{Camanho:2009vw,Chowdhury:2012km, Camanho:2014apa}) showed that in a theory satisfying an integrated null energy condition, the constraints are much stronger:
\be\label{hm}
n_s \geq 0   , \quad n_f \geq 0 \ , \quad \tn_s \geq 0 , \quad \tn_f \geq 0 , \quad \tn_v 
\geq  0 \ .
\ee
In large-$N$ holographic theories, identical constraints have been derived from causality on the gravity side \cite{Brigante:2008gz,Camanho:2009vw,Hofman:2009ug,Buchel:2009sk}. 

More generally, however, it is an open question whether these constraints hold in every unitary CFT, and how they are related to causality.\footnote{A partial argument based on the OPE of non-local energy expectation values appears in \cite{Hofman:2009ug}. A rather different argument, based on finite-temperature correlators, intriguingly leads to the same constraints \cite{Kulaxizi:2010jt} but is also incomplete as it requires the use of the Euclidean OPE beyond its usual regime of validity. A recent discussion suggesting that the average null energy condition might be violated in acceptable theories can be found \cite{Farnsworth:2015hum}. Significant progress on this question was also made very recently in \cite{Komargodski:2016gci}, where it was shown that the Hofman-Maldacena constraints follow from the optical theorem in a deep inelastic scattering experiment, assuming that the amplitude obeys a strong enough growth condition at large $|s|$.}  We will partially resolve this. Taking various polarizations in \eqref{nulle} leads to
\be\label{ourCA}
n_s\geq0 \ , \qquad n_s + 8 (d-1) n_f \geq 0 
\ee
and
\be
\tn_s \geq 0 , \qquad (d-2)^2\tn_s + 2(d-1)^3 \tn_f \geq 0 , \qquad \  \frac{d(d-2)^2}{(d-1)^3} \tn_s + 4 d \tn_f + 128 \tn_v \geq 0 .\label{ourCB}
\ee
In three dimensions, there is no vector structure, so only the first two inequalities apply.
In four dimensions, \eqref{ourCB} implies that the anomaly coefficients $a$ and $c$ satisfy
\be\label{acrange}
\frac{13}{54} \leq \frac{a}{c} \leq \frac{31}{18} \ .
\ee

\subsection{Discussion}
The constraints \eqref{ourCA} and \eqref{ourCB} are stronger than positivity of the 2-point function \eqref{wardc} but weaker than (and implied by) the Hofman-Maldacena energy conditions. They are necessary for causality, but we do \textit{not} claim they are sufficient.  Our derivation seems to exploit only a limited form of reflection positivity, so a better organization of positivity may lead to stronger constraints. 

One approach to strengthen the causality constraints may be to combine the methods here with the lightcone bootstrap \cite {Komargodski:2012ek,Fitzpatrick:2012yx,Fitzpatrick:2014vua,Alday:2014tsa,Vos:2014pqa,Fitzpatrick:2015qma,Alday:2015ota,Kaviraj:2015xsa,Kaviraj:2015cxa,Alday:2015ewa,Li:2015itl}. Causality fixes the sign of the anomalous dimensions of certain high-spin operators, and, barring some unexpected cancellations, we expect this statement to actually be stronger than the inequalities above. The anomalous dimensions can be computed by the lightcone bootstrap, but the calculation has not yet been done (in the required regime). Further remarks on the connection to the lightcone bootstrap are in section \ref{ss:anom}. It would also be very interesting to better understand the relationship between the present work and other recent work on related constraints, such as \cite{Maldacena:2015iua,Komargodski:2016gci,Fitzpatrick:2016thx}.

One motivation for this work was the recent demonstration, using causality on the gravity side, that holographic CFTs have $a \approx c$ \cite{Camanho:2014apa} (see also \cite{Papallo:2015rna,Bellazzini:2015cra}). As discussed at length in the introduction to \cite{paper1}, the weaker (but universal) constraint \eqref{acrange} may be a step toward deriving this surprising feature of large-$N$ sparse CFTs directly from the conformal bootstrap.

\subsection{Overview}

In the next section, we calculate the leading corrections to correlators across a shockwave. The result is proportional to \eqref{nulle}, and specializing to $O=J$ and $O=T$ with various polarizations leads to the combinations in \eqref{ourCA} and \eqref{ourCB}. Requiring these corrections to be negative gives the constraints. 

In the rest of the paper (which is almost independent from section \ref{s:shock}) we show \textit{why} these corrections must be negative. First we derive positivity conditions for spinning correlators, from reflection positivity, in section \ref{s:rp}. This essentially reduces the problem of spinning causality constraints to the scalar case.  The derivation of the causality constraints is in section \ref{s:causality}. We conclude with a discussion of the relationship to the lightcone bootstrap and anomalous dimensions in section \ref{ss:anom}.

\section{Correlators across a shockwave}\label{s:shock}

In this section, we summarize how the constraints in the introduction follow naturally from computing correlators in a `shockwave state',
\be
|\Psi\rangle = \psi(t=-i\delta)|0\rangle \ .
\ee  
Viewed on scales much larger than $\delta$, the energy in this state $\langle \Psi | T_{\mu\nu} | \Psi \rangle$ is supported on an expanding null shell centered at the origin.\footnote{For other work on shockwaves in CFT, especially in relation to the AdS/CFT correspondence, see \cite{Cornalba:2006xk,Cornalba:2006xm,Cornalba:2007zb}. The shockwave state considered there is apparently different, though the analytic properties of the correlator are similar.} Consider the 4-point function
\be
\langle  \Psi| \, \varepsilon \sdot O(x_2) \, \varepsilon^* \sdot O(x_3)\,  | \Psi \rangle
\ee
where
\be
\varepsilon \sdot O = \varepsilon_{\mu\nu\dots}O^{\mu\nu\dots} \ .
\ee
In the limit where the probe operators are almost null separated $(x_2 - x_3)^2 \to 0$, the leading correction comes from stress tensor exchange,
\be
OO \to T \to \psi \psi   \ .
\ee
This interaction is controlled by the 3-point couplings in $\langle OOT \rangle$, and by the scaling dimension of $\psi$, which fixes the coupling $\langle \psi\psi T\rangle$ via the Ward identity,
\be
c_{\psi\psi T} = -\frac{\Gamma(d/2)d}{2(d-1)\pi^{d/2}}\Delta_\psi \ .
\ee 
The magnitude of this correction \textit{grows} with the energy of the shock $E_{shock} \sim \frac{1}{\delta}$. For scalar probes, it was shown in \cite{paper1} that such corrections must be negative, or causality is violated. Below, we will show that the same is true for spinning probes: unitary theories must obey
\be\label{guess}
\frac{\langle \psi\,  \varepsilon \sdot O \, \varepsilon^* \sdot O \psi\rangle }{\langle \varepsilon \sdot O\,  \varepsilon^* \sdot O \rangle \langle \psi\psi\rangle} \approx 1 - {(\rm positive)} \frac{|x_2-x_3|^{d/2-1}}{\delta}
\ee
in the limit $0<(x_2-x_3)^2 \ll \delta \ll 1$. The basic reason is that if the correction has the wrong sign, then analyticity forces the correlator to have a singularity before the Minkowski lightcone, and this singularity produces a non-zero commutator at spacelike separation. For now, we just assume \eqref{guess} and work out the resulting constraints. In a theory with a low-dimension scalar appearing in the $OO$ OPE, $\frac{d}{2}-1 < \Delta < d-2$, the leading term is not of the form \eqref{guess} and is not constrained, so our analysis does not apply.

\subsection{$J$ and $T$ in $d=4$}

We start with a direct, full calculation of the stress tensor contribution to the shockwave correlator for conserved current probes in $d=4$. This approach is conceptually straightforward but requires a computer to do the algebra.  It also hides a dramatic simplification that occurs in the lightcone limit which we return to below.

The full conformal blocks for stress tensor exchange are calculated following \cite{cppr2}.  Details, and all of our conventions, are in the appendices. The output of this calculation is the spinning conformal block expressed as derivatives of the scalar conformal block.  The scalar block cannot be written in closed form in odd dimensions, but we will only require the leading terms in the lightcone limit, so this could be done in any $d$. 

\begin{figure}
\centering
\includegraphics{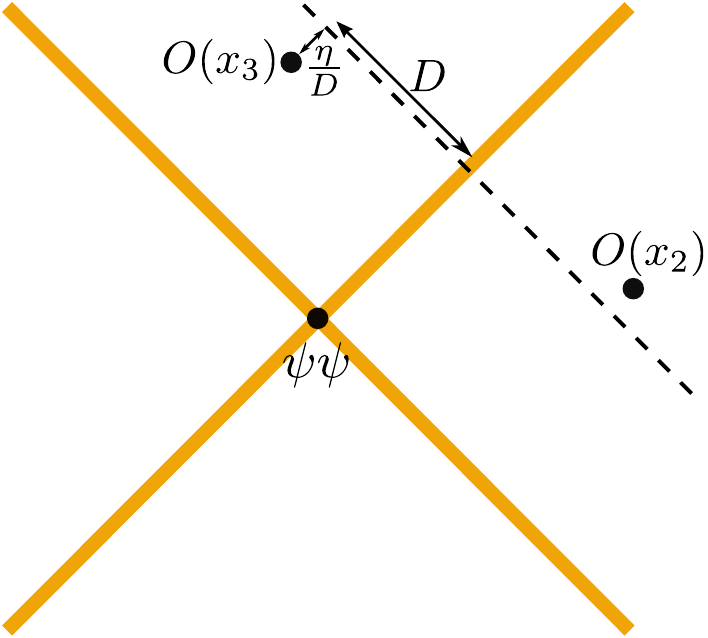}
\caption{\small Kinematics for the shockwave correlator. $O$ is a spinning probe in the state produce by the $\psi$ insertion. The two $O$'s are inserted symmetrically across the shock, and the two $\psi$'s are offset into the imaginary time direction (not drawn), $t = \pm i\delta$.\label{fig:kinematics}}
\end{figure}

\begin{figure}[h]
\centering
\includegraphics{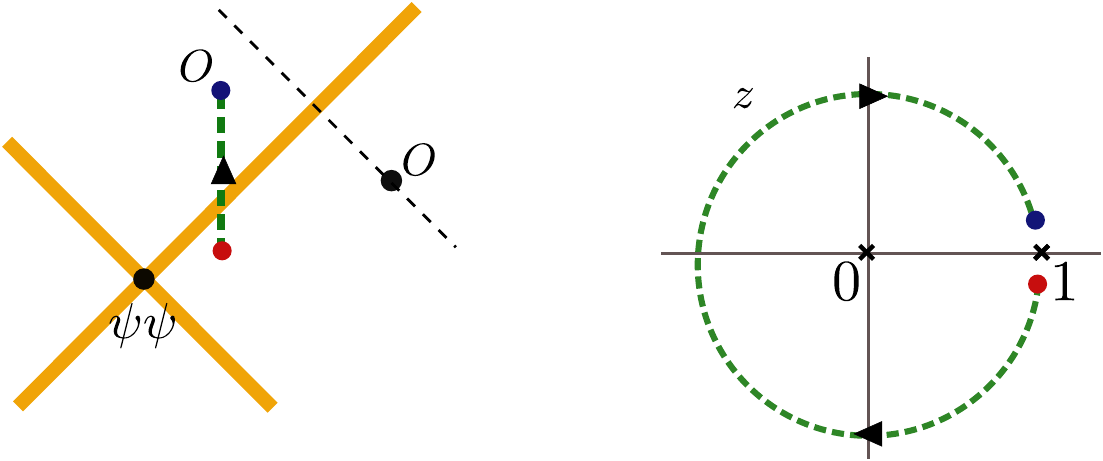}
\caption{\small The analytic continuation appropriate to the ordering $\psi OO \psi$ is shown here in real spacetime (\textit{left}) and on the $z$ plane (\textit{right}). As we move the $O$ operator into the future lightcone of the shockwave operator insertions, we cross a branch cut in the correlator. In terms of the cross-ratios, crossing this branch cut (with the correct $i\epsilon$ prescription for this ordering) leads to $z$ circling the origin.\label{fig:zcontour}}
\end{figure}

We work in $R^{1,3}$, with coordinates $(t,y,x_3,x_4)$ and place all insertions in the $ty$-plane. The operators creating the shockwave are inserted at
\be\label{xshockA}
x_1 = (t_1, y_1) = (i\delta,0)  , \quad
x_4 = (t_4, y_4) = (-i\delta, 0 ) \ ,
\ee
and the probes are inserted symmetrically across the shock:
\begin{align}\label{xshock}
x_2 &=\half\left(1-D+\frac{\eta}{D}, 1+D + \frac{\eta}{D}\right)\\
x_3 &=\half\left(1+D-\frac{\eta}{D}, 1-D - \frac{\eta}{D}\right) \notag \ .
\end{align}
We will also use the lightcone coordinates
\be
y^\pm = y \pm t \ .
\ee
$D$ is the distance to the shockwave, and the lightcone limit is $y_{23}^+ \sim  \frac{\eta}{D} \to 0$, as illustrated in figure \ref{fig:kinematics}.  
With these kinematics, we calculate the contribution from the identity and the stress tensor to the correlator
\be\label{gbbf}
G(\varepsilon, \eta, D) = \frac{\langle \psi(x_1)\,  \varepsilon \sdot O(x_2) \, \varepsilon^* \sdot O(x_3) \psi(x_4)\rangle }{\langle \varepsilon \sdot O(x_2)\,  \varepsilon^* \sdot O(x_3) \rangle \, \langle \psi(x_1)\psi(x_4)\rangle} \ ,
\ee
 in the limit $\eta \ll \delta \ll 1$, and compare to \eqref{guess}. The conformal cross ratios in the configuration (\ref{xshockA}-\ref{xshock}) are
 \be
 z=\frac{(D+i \delta )^2}{(D-i \delta )^2}\ , \qquad \bz=\frac{(D+i \delta  D-\eta ) (D-i \delta  D+\eta )}{(D-i \delta  D-\eta ) (D+i \delta  D+\eta )}\ .
 \ee
 In the limit $\eta \ll \delta \ll 1$, the cross-ratios are both near 1: 
\be\label{crossshock}
 z = 1+\frac{4i \delta}{D}, \qquad \bz = 1 + \frac{4i \delta}{D}\eta \ .
\ee
An important subtlety is that the conformal blocks are multivalued, and must be evaluated on the correct sheet. The prescription for the shockwave correlator, with operators ordered as in \eqref{guess}, was worked out in \cite{paper1}. The prescription to deal with branch cuts amounts to sending $z \to z e^{-2\pi i}$ (with $\bz$ held fixed) before evaluating the conformal blocks (see figure \ref{fig:zcontour}).

\subsubsection{Results for $\langle \psi JJ \psi\rangle$}\label{ss:jjres}
For a spin-1 probe $O=J_\mu$ in $d=4$, the two-point function in the denominator is
\be
\langle \varepsilon \sdot J(x_2)\,  \varepsilon^* \sdot J(x_3) \rangle  = -c_J\frac{D^2}{128\eta^4}\left[|\e^+|^2 - \frac{2\eta}{D^2} |\e^\perp|^2 + \frac{\eta^2}{D^4}|\e^-|^2\right] \ ,
\ee
where $\e^\perp$ is in the 3,4-directions. It is clear from this expression that different polarizations will contribute at different orders in the lightcone limit $\eta \to 0$. For any given polarization, the stress tensor corrections to the four-point function must be suppressed compared to the identity, so it will have a similar polarization dependence in the $\eta$ expansion.

Adding the stress tensor conformal block to the identity contribution and evaluating this in the lightcone limit, for a generic polarization the leading term in \eqref{gbbf} comes from the $J_+J_+$ polarization and is
\be
G(\varepsilon,\eta,D) = 1 -  \frac{5 D\Delta_\psi}{\pi c_J c_T }\cdot\frac{\eta}{\delta}n_s \ .
\ee
Therefore $n_s\geq 0$. If we instead choose a polarization with $\e^+=0$, this eliminates the leading terms in both the numerator and denominator of \eqref{gbbf}, giving
\be
G(\varepsilon,\eta,D) = 1 - \frac{5 D\Delta_\psi}{2\pi c_J c_T}\cdot\frac{\eta}{\delta}(n_s+24n_f)\ .
\ee
So $n_s+24n_f \geq 0$. There is no new constraint from the sub-sub-leading terms.

\subsubsection{Results for $\langle \psi TT \psi\rangle$}\label{ss:ttres}
For the stress tensor in $d=4$, choose without loss of generality a symmetric polarization tensor with $\varepsilon_{34} = \varepsilon^{\mu}_{\ \mu} = 0$. For the configuration (\ref{xshockA}-\ref{xshock}), the two-point function of the stress tensor is
\begin{align}
\langle \varepsilon \sdot T(x_2)\,  \varepsilon^* \sdot T(x_3) \rangle  = c_T\frac{D^4}{1024\eta^6}\left[|\e^{++}|^2 - \frac{4\eta}{D^2}|\e^{+\perp}|^2+ \frac{2\eta^2}{D^4}\left(|\e^{-+}|^2+2|\e^{\perp\perp}|^2\right)\right. \nonumber \\
\left.- \frac{4\eta^3}{D^6}|\e^{-\perp}|^2+\frac{\eta^4}{D^8}|\e^{++}|^2 \right] \ ,
\end{align}
where, $|\e^{\pm\perp}|^2=|\e^{\pm 3}|^2+|\e^{\pm 4}|^2$ and $|\e^{\perp\perp}|^2=|\e^{33}|^2+|\e^{44}|^2$. For any given polarization, the stress tensor correction to the correlator $\langle \psi TT \psi\rangle$ is suppressed compared to the identity, so it has a similar polarization dependence in the $\eta$ expansion. A generic choice gives
\be
G(\varepsilon,\eta,D) = 1 -   \frac{2560D \Delta_\psi}{3\pi c_T^{2}} \cdot \frac{\eta}{\delta} \tn_s \ ,
\ee
coming entirely from the $T_{++}T_{++}$ polarization. So $\tn_s\geq 0$. Choosing $\varepsilon^{++} = 0$ once again eliminates the leading terms in numerator and denominator, giving 
\be
G(\varepsilon,\eta,D) = 1 - \frac{640D \Delta_\psi}{3\pi c_T^{2}} \cdot\frac{\eta}{\delta}(2\tn_s + 27 \tn_f) \ ,
\ee
the second constraint: $2\tn_s + 27 \tn_f\ge 0$.
Now choosing $\varepsilon^{+3} = \varepsilon^{+4} = 0$ eliminates this term as well, giving
\be\small
G(\varepsilon,\eta,D) = 1 - \frac{1280D \Delta_\psi \eta}{9\pi c_T^{2}\delta}\left[ \frac{\left(\tn_s+27\tn_f+216\tn_v\right)|\e^{33}-\e^{44}|^2+4(8\tn_s + 27 \tn_f)|\e^{33}+\e^{44}|^2 }{|\e^{33}-\e^{44}|^2+2|\e^{33}+\e^{44}|^2}\right]
\ee
Both coefficients must be positive, giving us the third constraint: $\tn_s+27\tn_f+216\tn_v>0$. (The inequality $8\tn_s + 27 \tn_f\ge 0$ is not independent, and no new constraints come from the remaining polarizations.)

Said differently, the three constraints come from (for example)
\begin{align}
\langle \psi [T_{++}, T_{++}] \psi\rangle = 0 \ , \quad \langle \psi [T_{+3}, T_{+3}]\psi\rangle = 0 , \quad \langle\psi [T_{33}-T_{44}, T_{33}-T_{44}]\psi\rangle = 0 \ ,
\end{align}
respectively.

\subsection{General probes}
Now consider a spin-$\ell$ probe, not necessarily conserved, in any spacetime dimension $d$.  In the previous examples, we computed the full conformal blocks first, then took the lightcone limit. This is not really necessary, and is not practical for general probes, so here we will use a more efficient approach in which we compute the lightcone blocks directly.  (Of course we could have done this for $J$ and $T$ as well, but the full blocks are useful for other purposes and provide a nontrivial check.)

\subsubsection{Spinning lightcone blocks}

Define the partial waves for $X$ exchange by the OPE in the $O\to O$ channel,
\be
\langle  \e_1\sdot O(0)  \e_2 \sdot O(y^+, y^-) \psi(1\hat{y}) \psi(\infty)\rangle = \sum_X W_X(\e_1, \e_2, z, \bz) \ .
\ee
The cross ratios in this configuration are simply
\be
z = y^-  , \qquad \bz = y^+  \ .
\ee
The lightcone blocks are defined as the leading contribution in the limit $\bz \to 0$ with $z$ held fixed. The components of the polarization tensors $\e_i$ are also held fixed, so there are relations among the tensor structures in this limit. 

The conformal block greatly simplifies in the lightcone limit, even for external operators with spin. The simplification comes from reorganizing the conformal families under the colinear $SL(2)$ that preserves the light ray so that the problem becomes effectively two-dimensional, as described in \cite{Komargodski:2012ek,Fitzpatrick:2015qma}. Take $X$ to be an arbitrary symmetric traceless tensor. The full $d$-dimensional conformal block is built from 3-point functions with $X$ and its conformal descendants, 
\be\label{wxsum}
W_X = \sum_{D}\langle \e_1 \sdot O(0) \e_2 \sdot O(z,\bz)  D X_{\mu\nu\dots}(\infty \hat{y})\rangle \langle D^{\dagger}X_{\alpha\beta\dots}(0) \psi(1) \psi(\infty)\rangle K^{\mu\nu\dots, \alpha\beta\dots} \ , \ee
where $D$ is a differential operator built from the conformal generators and $K = \langle D^{\dagger}X|DX\rangle^{-1}$.  The dominant contributions to \eqref{wxsum} in the lightcone limit come from the component $X_{--\cdots-}$ and its $\p_-$ derivatives. The $\p_-$ derivatives can be summed as in 2d CFT.\footnote{For spinning operators, it matters what direction we define to be `$\infty$' in $X_{\mu\nu\dots}(\infty)$ and the decomposition \eqref{wxsum} depends on this choice, although of course the final sum does not. We take $X$ to infinity in the $y$-direction, $X(\infty) \equiv \lim_{r\to \infty}r^{\Delta_X}X(r\hat{y})$. A choice outside of the $ty$-plane would also work but requires swapping $X_{--\dots-} \to X_{++\dots +}$ throughout the following discussion, including in the final energy condition \eqref{nullebody}.}

In the end, as anticipated in \cite{Komargodski:2012ek,Fitzpatrick:2015qma}, we find a remarkably simple formula for the spinning lightcone block (or rather partial wave, since all the coefficients and prefactors are included):
\be\label{wlc}
W^{LC}_X(\e_1,\e_2,z,\bz) = \frac{c_{\psi\psi X}}{c_X}\langle \e_1 \sdot O(0) \e_2 \sdot O(z,\bz) X_{--\dots-}(\infty\hat{y})\rangle \, _2F_1(h_X,h_X,2h_X,z)\ ,
\ee
where $h_X = \half(\Delta_X+\ell_X)$ and $c_X$ is the normalization constant of the two-point function.  For any given choice of external polarizations, this formula gives the leading non-zero term in $W_X$ as $\bz  = y^+\to 0$.\footnote{What power of $\bz$ this is depends on the polarizations. To extract the full content of \eqref{wlc} one must first choose a definite polarization, and then compute the leading non-zero term.  If we compare the $\bz$-expansion of $W_X$ to the $\bz$ expansion of the right-hand side of \eqref{wlc}, with arbitrary $\e$'s, the leading-$\bz$ term on both sides comes from $O_{++\dots+}$ polarizations, and matches exactly. The subleading-$\bz$ terms come from $O_{++\dots+3}$, and these terms match \textit{only} if we choose the polarization such that the leading all-$+$ terms vanish, and so on for sub-sub-leading terms. Also note that we have assumed $\ell_X$ is even since this is required for $c_{\psi\psi X} \neq 0$.} 

This simplification is by no means apparent from the full block, calculated using \cite{cppr2}, which is expressed as derivatives on scalar blocks of various weights. Nonetheless we have compared the two for $\ell_O = 1,2,3$ (and arbitrary $\Delta_O,\Delta_X$, $\ell_X$), and confirmed that the two methods agree after applying a variety of hypergeometric identities.

\subsubsection{Constraint on coupling to null energy}
With the formula for the lightcone block \eqref{wlc} in hand we now return to the causality constraints. 
By a conformal transformation, the sign constraint \eqref{guess} can be rephrased in terms of the stress tensor partial wave as
\be\label{ttpot}
\frac{W_T(\e, \e^*, z, \bz)}{\langle  \e\sdot O(0)  \e^* \sdot O(z,\bz) \rangle} \approx \mbox{(positive)}\times \frac{i \bz^{d/2-1}}{z^{d/2}}  \ ,
\ee
in the limit $|\bz| \ll |z| \ll 1$. (We have transformed $z\to 1-z$, $\bz \to 1-\bz$ compared to \eqref{crossshock} for notational convenience.) The hypergeometric function in \eqref{wlc} has a branch cut along $z \in (1,\infty)$, and we must go around the cut to calculate the shockwave correlator in the correct operator ordering.  This sends
\be
_2F_1(h,h,2h,z) \to\  _2F_1(h,h,2h,z) + 2 \pi i \frac{\Gamma(2h)}{\Gamma(h)^2} \ _2F_1(h,h,1,1-z) \ .
\ee
Expanding now for $z \sim 0$ and using $h_T = d/2+1$, \eqref{wlc} gives
\begin{align}
W^{LC}_T(\e, \e^*, z, \bz) &\approx  -i \frac{2\Delta_\psi(d+1)\Gamma(d+1)^2}{ \pi^{\frac{d}{2}-1}c_T(d-1)\Gamma(\half(d+2))^3} \\
& \qquad \quad \times z^{-1-d} \langle \e_1 \sdot O(0) \e_2 \sdot O(z,\bz) T_{--}(\infty \hat{y})\rangle \ .\notag
\end{align}
Comparing to the positivity condition \eqref{ttpot} gives the result stated in the introduction,
\be\label{nullebody}
\frac{ \langle \varepsilon\sdot O(0) \varepsilon^* \sdot O(y^+,y^-) T_{--}(\infty \hat{y})\rangle }{\langle \varepsilon\sdot O(0) \varepsilon^* \sdot O(y^+,y^-) \rangle} < 0
\quad \mbox{as} \quad y^+ \to 0 \ ,
\ee
for spacelike separated probes, $y^+y^-> 0$.

This is easily calculated for the examples $O=J$ and $O=T$ in $d=4$ using the 3-point functions in appendix \ref{s:currents}. The results agree with the brute force calculations above. Obviously this method is much easier than the shockwave calculation, since we only need to evaluate a 3-point function. Furthermore, the inequality (\ref{nullebody}) can be used to extend our results to arbitrary dimensions which leads to constraints (\ref{ourCA})  and (\ref{ourCB}).

\subsubsection{Example: Non-conserved spin 1}
It is also straightforward now to derive constraints on other operators. Consider, for example, a non-conserved spin-1 operator $V$.  The 3-point function $\langle VVT\rangle$ can be expanded in the basis \eqref{jjtbasis} used for $\langle JJT\rangle$, but now with fewer conditions on the $\alpha_i$. Conservation of the stress tensor imposes $\alpha_5 = \frac{2}{d^2-4}(\alpha_1 - 2 \alpha_3)$, leaving the coupling constants $\alpha_{1,2,3}$. Evaluating \eqref{nullebody} for the polarizations $+$, $3$, and $-$, respectively, leads to the constraints
\be
\alpha_1 - 2 \alpha_2 \geq 0\  , \quad -\alpha_2 \geq 0 \ , \quad (d^2+4)\alpha_1 + 2(4-d^2)\alpha_2 - 4 d^2 \alpha_3 \geq 0 \ .
\ee

\section{Reflection Positivity for Spinning Correlators}\label{s:rp}

The goal in the rest of the paper is to derive the statement that the corrections in \eqref{guess} must be negative. The strategy is very similar to our analysis of scalar probes \cite{paper1}. The basic idea is to show that a small positive correction in the shockwave kinematics would imply that the four-point function is actually singular \textit{before} the expected Minkowski lightcone, thereby violating causality.

An essential step in the argument is to bound the analytically continued correlator on the 2nd sheet (after taking $z \to z e^{-2\pi i}$) by its value on the 1st sheet. This relies on finding a function $F_{\varepsilon}(z,\bz)$, constructed from the spinning correlator, that has an expansion near $z,\bz \sim 0$ with positive coefficients.  It should be a function of the cross ratios that may depend on a choice of polarization, but cannot depend explicitly on the $x_i$ (unlike the correlator itself, which has explicit $x_i$-dependence in the tensor structures). For external scalars \cite{paper1}, this function with a positive expansion was the correlator itself, $F_{scalar}(z,\bz) = \langle \psi(0) \phi(z,\bz) \phi(1) \psi(\infty)\rangle$. However, the spinning correlator   $\langle \psi(0) \, \varepsilon \sdot O(z,\bz) \, \varepsilon^* \sdot O(1) \psi(\infty)\rangle$ does \textit{not} have an expansion with any fixed sign, so we must work harder to construct such a function, starting from reflection positivity of the spinning correlator. This is the goal of this section.

Although we will give a prescription to construct a positive $F$, we do not claim that this construction is optimal, or that the resulting positivity conditions will exploit the full consequences of reflection positivity.  This construction will be enough to prove the sign constraints in the shockwave discussed in section \ref{s:shock}, but it seems likely that a more clever use of reflection positivity would produce stronger constraints (see more comments on this in section \ref{ss:anom} below).

Note that in the rest of this section, we work in Euclidean signature, whereas section \ref{s:shock} was in Lorentzian notation.

\subsection{Review of reflection positivity}

Reflection positivity is the statement that certain correlators in a Euclidean QFT are real and positive, and is required for a Euclidean theory to have a unitary Lorentzian counterpart. For example, for real scalars, if we choose (arbitrarily) one of the coordinates of $R^d$ to call Euclidean time $\tau$, and label points by $x = (\vec{x}, \tau)$, then reflection positivity implies
\be\label{scalarpos}
\langle \phi(\vec{x}_1, \tau_1) \phi(\vec{x}_2, \tau_2) \cdots \phi(\vec{x}_n, \tau_n) \phi(\vec{x}_n, -\tau_n) \cdots \phi(\vec{x}_2, -\tau_2)\phi(\vec{x}_1, -\tau_1)\rangle  > 0 \ .
\ee
In Hilbert space language, this ensures that the state on the plane at $\tau=0$ created by doing the path integral over the lower half plane with the insertions at $x_{1,\cdots, n}$ has positive norm.  It follows automatically from the path integral, assuming a real Lagrangian. For example setting $n=2$, the path integral for \eqref{scalarpos} is manifestly positive:
\begin{align}
\int_{R^d} D\phi \, & \phi(\vec{x}_1, \tau_1) \phi(\vec{x}_2, \tau_2) \phi(\vec{x}_2, -\tau_2)\phi(\vec{x}_1, -\tau_1)e^{-\int \mathcal{L}(\phi)} = \\
\ & \quad \int D \phi_0(\vec{x})\left| \int_{\phi(\vec{x}, 0)=\phi_0(\vec{x})}D\phi_{\tau>0} \,  \phi(\vec{x}_1, \tau_1) \phi(\vec{x}_2, \tau_2) e^{-\int_{\tau>0} \mathcal{L}(\phi) } \right|^2 \ \notag \ .
\end{align}
Reflection positivity also holds for sums or integrals of local operators, for the same reasons. For example,
\be\label{sumphi}
\langle \left( a \phi(\vec{x}_1, \tau_1) + b \phi(\vec{x}_2, \tau_2)\right) \left( a^* \phi(\vec{x}_1, -\tau_1) + b^* \phi(\vec{x}_2, -\tau_2)\right)\rangle  > 0 \ .
\ee
For operators with spin, reflection positivity is just how it sounds: configurations which are symmetric under reflection $\tau \to -\tau$ must have positive correlators. Reflection acts on vector indices as well:
\be
J_\mu = (J_\tau(\vec{x}, \tau), J_y(\vec{x}, \tau), \dots ) \mapsto \left[J_\mu(\vec{x}, \tau)\right]^{R} \equiv \left(-J_\tau(\vec{x}, -\tau), J_y(\vec{x}, -\tau), \dots \right)  \ ,
\ee
and the positive combination is
\be
\langle \dots J_\mu(\vec{x}, \tau) [J_\mu(\vec{x}, \tau)]^R \dots \rangle > 0 \ ,
\ee
assuming the dots are also reflection-symmetric.  This can be understood from the path integral, or by interpreting \eqref{sumphi} for nearby points as a derivative. For higher spin, all of the $\tau$ indices are reflected, with $\vec{x}$ indices unaffected. To write this succinctly, we define the reflected polarization tensor
\be
\varepsilon^R_{\mu\nu\dots} = (-1)^{\mbox{\footnotesize \# of $\tau$ indices}} \varepsilon^*_{\mu\nu\dots} \ ,
\ee
so that
\be
\langle \dots \varepsilon \sdot O(\vec{x}, \tau)\,  \varepsilon^R \sdot O(\vec{x}, -\tau) \dots \rangle > 0  \ .
\ee
In Lorentzian signature, reflection simply complex-conjugates the components of $\varepsilon$ with no need for the factors of $-1$.

\subsection{Positive expansions for spinning correlators}\label{ss:pos}
Reflection positivity implies that a scalar correlator $\langle \psi(0) O(z,\bz) O(1) \psi(\infty)\rangle$ has an expansion around $z,\bz\sim 0$ with positive coefficients \cite{Fitzpatrick:2012yx, Hogervorst:2013sma, paper1}. The derivation of this statement in appendix A of \cite{paper1} can be adapted to find analogous results for spinning probes.  Define
\be\label{mg}
G_{\varepsilon}(\tau_1, y_1; \tau_2, y_2) = \langle \psi(1,0) \varepsilon \sdot O(\tau_1, y_1) \varepsilon^R \sdot O(-\tau_2, y_2) \psi(-1,0)\rangle \ .
\ee
Operators are inserted in the $(\tau,y)$ plane, with other positions $\vec{x}_i=0$ not written. $\varepsilon$ is a constant polarization tensor. Reflection positivity requires
\be\label{intposg}
\int_0^{\infty}d\tau_1 d\tau_2 \int_{-\infty}^{\infty} dy_1 dy_2 \, \Omega(y_1, \tau_1) \Omega^*(y_2, \tau_2) G_\varepsilon(y_1, \tau_1; y_2, \tau_2) > 0 \ ,
\ee
for any smearing function $\Omega$. Let us change integration variables to map the upper half plane to the unit disk,
\be\label{kksl}
r_1 e^{i\theta_1} = \frac{y_1 + i \tau_1 - i}{y_1 + i \tau_1 + i}\ , \qquad r_2 e^{i\theta_2} = \frac{y_2 + i \tau_2 - i}{y_2 + i \tau_2 + i}  \ .
\ee
The cross ratios for the configuration \eqref{mg}, with points labeled as
\be
\psi(x_1) O(x_2) O(x_3) \psi(x_4)
\ee
are 
\be
z = r_1 r_2 e^{i(\theta_1- \theta_2)} , \qquad \bz = r_1 r_2 e^{-i(\theta_1-\theta_2)} \ .
\ee
Now we plug in the general form of a spinning correlator \eqref{genfour}, and absorb terms that factorize as $g(r_1,\theta_1)\times g(r_2,\theta_2)$ into the definition of $\Omega$. The positive quantity \eqref{intposg} becomes
\begin{align}\label{posbec}
\int_0^1 dr_1 dr_2 \int_{0}^{2\pi} d\theta_1 d\theta_2 \, &\Omega(r_1,\theta_1)\Omega^*(r_2,\theta_2)\sum_A f_A(z,\bz) Q^A(r_1,r_2,\theta_1,\theta_2; \varepsilon) \ ,
\end{align}
where the sum is over tensor structures $Q^A$. 

Before we proceed, note that these $Q^A$'s cannot be written as functions of $z$,$\bz$.  However, we can project them onto functions of $z$ and $\bz$, with an operation we will denote as
\be
Q^A(z,\bz;\varepsilon) = \left[Q^A(r_1, r_2,\theta_1,\theta_2; \varepsilon)\right]_{proj} \ . 
\ee
The projection is defined by expanding $Q^A$ in powers of $r_1, r_2, e^{i\theta_1}, e^{i \theta_2}$, and keeping only the terms with the same power of $r_1$ and $r_2$, and opposite powers of $e^{i \theta_1}$, $e^{i\theta_2}$:
\be
\left[ r_1^m r_2^n e^{i k \theta_1 - \ell \theta_2} \right]_{proj} \equiv z^{\half(m+k)}\bz^{\half(m-k)}\delta_{mn}\delta_{k\ell} \ .
\ee
Some examples are worked out below.

Now return to \eqref{posbec}. As explained in \cite{paper1}, we can choose $\Omega(r_1,\theta_1)$ to project onto particular  powers of $r_1$ and $e^{i\theta_1}$. Then $\Omega^*(r_2, \theta_2)$ will project onto the same power of $r_2$, and opposite power of $e^{i\theta_2}$. Therefore, with this choice of smearing function, the integral \eqref{posbec} is completely insensitive to the difference $Q^A - [Q^A]_{proj}$.  That is, we can replace $Q_A \to [Q_A]_{proj}$ under the integrand. Then following \cite{paper1} we conclude that the function
\be
F_\varepsilon(z,\bz) \equiv \sum_A f_A(z,\bz) Q^A(z,\bz; \varepsilon)
\ee
has an expansion in $z,\bz$ with positive coefficients:
\be\label{fposre}
F_\varepsilon(z,\bz)=\sum_{h,\bh}(\mbox{positive})\times z^h \bz^{\bh} \ .
\ee

\subsection{Examples}\label{ss:examples}
We will see below that to derive constraints, we never need to actually perform the projection.  It is enough that it exists.  But to clarify the procedure we will work out some applications to $\langle JJ\psi\psi\rangle$, which may be of interest on their own. This subsection is unnecessary for the rest of the paper.

Conformal invariance fixes the form
\begin{align}\label{jstruct}
\langle \psi(x_1)&  \varepsilon_2 \sdot J(x_2) \varepsilon_3 \sdot J(x_3) \psi(x_4)\rangle = \\
& \left( \frac{x_{24}x_{13}}{x_{14}^2}\right)^{\Delta_\psi-d}\left(x_{12}x_{34}\right)^{\Delta_\psi+d}  \left(f_1 H_{23} + f_2 V_2 V_3 + f_3 V_2' V_3 + f_4 V_2 V_3' + f_5 V_2' V_3'\right)\notag
\end{align}
where the tensor structures $V_i$ and $H_{ij}$ are standard notation defined in appendix \ref{s:structure}, and the coefficient functions $f_A = f_A(z,\bz)$ are arbitrary functions of the cross ratios. Let us choose a transverse polarization,
\be
\varepsilon = \varepsilon^R = \hat{x}_3 \ ,
\ee
where $x_3$ is one of the directions other than $(\tau,y)$. In the configuration \eqref{mg}, the only non-zero tensor structure is
\bea
H_{23} &=& (y_1-y_2)^2  + (\tau_1+\tau_2)^2\\
&=& 4\frac{1 + r_1^2r_2^2 - 2 r_1 r_2 \cos(\theta_1-\theta_2)}{(1+r_1^2 + 2r_1 \cos\theta_1)(1 + r_2^2 + 2 r_2 \cos\theta_2)}\ . \notag
\eea
Expanding out the denominator, only the first term survives the projection,
\be
\big[ H_{23} \big]_{proj} = 4 \ .
\ee
Therefore, 
\be\label{ftrans}
F_{transverse}(z,\bz)  =4 f_1(z,\bz)
\ee
has a positive expansion in $z,\bz$.

Applying the same logic to the polarization
\be
\varepsilon = \varepsilon^R = \hat{y} + i \hat{\tau}
\ee
we find, for example,
\be
\big[H_{23}\big]_{proj} = 8 \frac{1+z}{1-z} , \quad \big[V_2V_3\big]_{proj} = -4 \frac{1+z}{1-z} , \qquad \mbox{etc.},
\ee
leading to
\be\label{jpos2}
F_{parallel}(z,\bz) =4 \left(\frac{1+z}{1-z}\right)\left[2 f_1 - f_2 +\frac{1}{1-z} f_3 -z f_4 + \frac{z}{1-z} f_5\right] \ .
\ee
Choosing $\varepsilon \to \varepsilon^*$ gives a similar expression with $z \leftrightarrow \bz$. Thus in a reflection positive CFT, these combinations of coefficient functions have expansions in $z,\bz$ with positive coefficients. 

The statement that \eqref{ftrans} and \eqref{jpos2} have positive coefficients must hold also for individual conformal blocks in the channel
\be
J \psi \to X \to J\psi \ .
\ee
As a check, we computed these conformal blocks following \cite{cppr2} with $\ell_X=1,2,3,4$, and confirmed to high order that the expansion coefficients of both \eqref{ftrans} and \eqref{jpos2} are indeed positive.

Although we have only discussed constant $\varepsilon$, the procedure also applies to spacetime dependent polarizations.  We have not explored this possibility in much detail, but some simple choices did not yield any new constraints.

\section{Derivation of causality constraints}\label{s:causality}

Now that we have constructed a function $F_{\varepsilon}(z,\bz)$ with a positive expansion, we can immediately borrow the results of \cite{paper1} to derive causality constraints.  We simply treat $F_{\varepsilon}(z,\bz)$ as if it were a scalar correlator, and follow the same steps.\footnote{However we will skip one step to simplify the argument. In \cite{paper1}, we started by showing directly from the conformal block expansion (in the $\rho$-coordinate) that the correlator was analytic, as required for causality, then derived constraints.  The proof of analyticity was instructive, but is not necessary in order to derive constraints --- the Osterwalder-Schrader reconstruction theorem guarantees that any QFT obeying the usual Euclidean axioms is causal in Lorentzian signature \cite{os1,os2,haag}. Therefore in this paper we will skip the demonstration of analyticity, invoking the Osterwalder-Schrader theorem, and proceed directly to the constraints.} In this section we will review these steps to make the presentation self-contained, and show that this leads to the shockwave positivity condition \eqref{guess}.

\subsection{The sum rule}
The method hinges on the following observation  \cite{paper1} (which is related to earlier methods in \cite{Camanho:2014apa,Maldacena:2015waa}). Suppose the function $F_\eta(z)$, where $\eta$ is a positive real parameter, has the following properties near $z \sim 1$:
\begin{enumerate}
\item $F_{\eta}(z)$ is analytic in the upper half $z$-plane (not necessarily on  the real line); 
\item For real $x$ near 1, $F_\eta(x)$ is non-singular and Re $F_\eta(x) < 1$;
\item For $\eta \ll |1-z| \ll1 $, it has the expansion
\be\label{FFc}
F_\eta(z) = 1+ i \lambda \eta^a \left[ \frac{1}{1-z} + O(|1-z|^0)\right] + o(\eta^a) ,
\ee
where $a>0$.
\end{enumerate}
Then
\be
\lambda > 0 \ .
\ee
The proof is simple. Consider the semicircle contour 
\be
\includegraphics{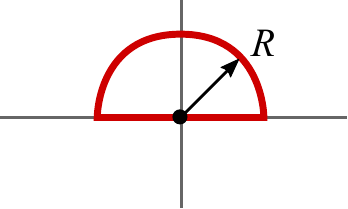}
\ee
where the origin is at $z=1$, and the radius satisfies $\eta \ll R \ll 1$. Analyticity implies $\mbox{Re} \oint dz F_\eta(z)=0$, which leads to the sum rule
\be\label{fsum}
\lambda = \lim_{R \to 0}\lim_{\eta \to 0}\frac{1}{\eta^a} \int_{-R}^R dx \, (1-\mbox{Re\ }F_{\eta}(x)) \ .
\ee
By assumption 2, this is positive.

\subsection{Positivity of shockwave corrections}

Starting with the function $F(z,\bz)$ constructed in section \ref{s:rp}, we can construct a function $F_\eta(z)$ obeying the assumptions of the previous subsection as
\be\label{fetarat}
F_\eta(z) \equiv \frac{F(ze^{-2\pi i}, \bz)}{F(z,\bz)} \ , \quad \mbox{where} \quad 1-\bz = \eta(1-z) \ .
\ee
Analyticity in the domain required by assumption 1 above follows from causality, as discussed in detail in \cite{paper1}. The fact that $F(z,\bz)$ has an expansion around $z,\bz \sim 0$ with positive coefficient \eqref{fposre} ensures Re $F_\eta(x) < 1$ on the real line, as required by assumption 2. (Actually this works only for $0<x<1$ where this expansion converges, but the same statement for $x>1$ follows from crossing.) 

The last step is to compute the leading terms from the lightcone OPE, and compare to the form in assumption 3. That is, we compute the leading terms in the correlator in the limit
\be\label{fnl}
|1-\bz| \ll |1-z| \ll 1 \ .
\ee
The denominator of \eqref{fetarat} has no corrections that grow as $(1-z)^{-1}$, so it can be replaced by the product of two-point functions.  Therefore, writing the numerator in the format \eqref{genfour}, 
\be
F_{\eta}(z) = \frac{\sum_A f_A(ze^{-2\pi i},\bz)[Q^A(x_i,\varepsilon)]_{proj}}{[H_{23}(x_i,\varepsilon)^{\ell_O}]_{proj}} \ ,
\ee
where $\bz = 1 + \eta z - \eta$. Ignoring the projections for a moment, this quantity in the limit \eqref{fnl} is exactly what was calculated in the context of shockwaves in section \ref{s:shock}, equation \eqref{gbbf}. In fact, this conclusion is unaffected by the projections: The leading terms computed in section \ref{s:shock} were proportional to $\frac{(1-\bz)^{d/2-1}}{(1-z)^{d/2+1}}$, which is invariant under the projection. 

Finally, comparing the corrections in \eqref{guess} to the form \eqref{FFc}, we see that the coefficient in \eqref{guess} obeys a sum rule and must be positive. This completes the derivation of the constraints listed in the introduction.

Another route to the same constraints is to work out the positive combinations of coefficient functions as in section \ref{ss:examples}. Then the scalar logic can be applied to each of these combinations.  As a check, we confirmed that this gives the same constraints for $\langle JJ \psi\psi\rangle$ in $d=4$. For example, if we choose the transverse polarization, then from \eqref{ftrans} the function with a positive expansion is $F(z,\bz) = f_1(z,\bz)$. The contribution to this coefficient function from stress tensor exchange $JJ \to T \to \psi\psi$ is given in \eqref{fj}. Evaluating this expression on the second sheet in the shockwave limit \eqref{fnl} gives a leading term proportional to $n_s + 24 n_f$, in agreement with the first constraint in \eqref{ourCA}. The other polarizations work similarly.

\section{Connection to the lightcone bootstrap}\label{ss:anom}

We conclude with some comments on the relationship to the lightcone bootstrap program \cite{Komargodski:2012ek,Fitzpatrick:2012yx,Fitzpatrick:2014vua,Alday:2014tsa,Vos:2014pqa,Fitzpatrick:2015qma,Alday:2015ota,Kaviraj:2015xsa,Kaviraj:2015cxa,Alday:2015ewa,Li:2015itl}, and speculate on how it might be used to strengthen the causality constraints. For concreteness, consider a spin-1 probe,
\be
\langle \psi(0) J(z,\bz) J(1) \psi(\infty)\rangle \ .
\ee 
Crossing symmetry is the statement that this correlator can be decomposed in two different OPE channels:
\\

\begin{equation}
\vspace{.5cm}
\sum_X \qquad
\begin{gathered}
\begin{fmffile}{schannel}
\begin{fmfgraph*}(40,20)
  	\fmfleft{i1,i2}
        \fmfright{o1,o2}
        \fmf{plain}{i1,v1,i2}
        \fmf{plain}{o1,v2,o2}
        \fmf{plain,label=$X$}{v1,v2}
	\fmflabel{$\psi$}{i1}
	\fmflabel{$J(z,\bz)$}{i2}
	\fmflabel{$J(1)$}{o2}
	\fmflabel{$\psi$}{o1}
\end{fmfgraph*}
\end{fmffile}
\end{gathered}
\qquad  =\qquad  \sum_{X'} \qquad 
\begin{gathered}
\begin{fmffile}{tchannel}
\begin{fmfgraph*}(40,20)
  	\fmfleft{i1,i2}
        \fmfright{o1,o2}
        \fmf{plain}{i1,v1,o1}
        \fmf{plain}{o2,v2,i2}
        \fmf{plain,label=$X'$}{v1,v2}
	\fmflabel{$\psi$}{i1}
	\fmflabel{$J(z,\bz)$}{i2}
	\fmflabel{$J(1)$}{o2}
	\fmflabel{$\psi$}{o1}
\end{fmfgraph*}
\end{fmffile}
\end{gathered}
\end{equation}
The lightcone bootstrap is a method to solve this equation in the lightcone limit,
\be
\bz \to 1  \qquad (\mbox{fixed\ } z) \ .
\ee
In this limit, the dominant contribution on the right-hand side of the crossing equation is the identity, and the first correction is the stress tensor.  On the left, these two terms are reproduced by summing over high-spin composite operators, schematically
\be\label{jptypes}
[J\psi]^{(S)}_{n,\ell} \sim J_{(\nu} \partial_{\mu_1)} \cdots \partial_{\mu_{\ell-1}}\Box^n \psi
\quad \mbox{and} \quad
[J\psi]^{(A)}_{n,\ell} \sim J_{[\nu} \partial_{\mu_1]} \cdots \partial_{\mu_{\ell-1}}\Box^n \psi
\ee
with $\ell \gg n \geq 0$. The identity term is produced by assigning these operators their canonical scaling dimensions and solving for the OPE coefficients, and the stress tensor correction comes from their anomalous dimensions $\gamma_{n,\ell}^{(A,S)}$.

To illustrate the various regimes, let us write the stress tensor lightcone block schematically (suppressing all tensor dependence) as
\be
W_T \sim (1-\bz)^{d/2-1}\left(f_1(z)\log(z)  + f_2(z) \right), 
\ee
where $f_{1,2}$ are meromorphic. The $n=0$ anomalous dimensions are determined by the leading behavior as $z \to 0$,
\be
\gamma_{0,\ell}\quad \leftrightarrow \quad |1-\bz| \ll |z|  \ll 1\quad  \leftrightarrow\quad  f_1(z \to 0) \ .
\ee
Anomalous dimensions for higher $n$ come from subleading terms in an expansion of $f_1(z)$ around $z\sim 0$. The asymptotics, $\ell \gg n \gg 1$, are determined by the behavior of $f_1(z)$ near $z \sim 1$ \cite{Kaviraj:2015cxa,Kaviraj:2015xsa}:
\be
\gamma_{\ell\gg n\gg 1}\quad  \leftrightarrow\quad  |1-\bz| \ll |1-z|  \ll 1 \quad \leftrightarrow\quad   f_1(z \to 1) \ .
\ee

The shockwave calculation relied on the stress tensor block in precisely this latter regime: we sent $z \to z e^{-2\pi i}$ and then evaluated the block for $|1-\bz| \ll |1-z| \ll 1$. The leading terms came from the non-analytic term, $f_1$ in the conformal block. The conclusion is that causality, in the form \eqref{guess}, also fixes 
\be
\gamma_{n,\ell}^{(A,S)} < 0 \qquad \mbox{for} \qquad \ell \gg n \gg 1 \ .
\ee
This constraint seems likely to be \textit{stronger} than the inequalities stated in the introduction. The reason is that the lightcone bootstrap effectively decomposes our positive function $F_\e(z,\bz)$ into two positive functions,
\be\label{fdecc}
F_{\e}(z,\bz) \sim F_{\e}^A(z,\bz) + F_{\e}^S(z,\bz)
\ee
coming from the two families of operators in \eqref{jptypes}. The entire causality analysis can then be repeated for these two functions individually. Unless the decomposition \eqref{fdecc} is degenerate for some reason (\ie only one of the two functions contributes to \eqref{fdecc} for any given polarization), the resulting constraints will be stronger. 

Of course to make this concrete the spinning lightcone bootstrap must be performed in the regime $\ell \gg n \gg 1$. So far, for external spinning operators it has only been done for $n=0$ in $d=3$ \cite{Li:2015itl}.  The result in that case is that the anomalous dimensions are proportional to the same combinations of couplings constrained by a Hofman-Maldacena type analysis,
\be
\gamma_{0,\ell}^{A} \propto -n_f \qquad \mbox{and} \quad \gamma_{0,\ell}^S \propto -n_s \ .
\ee
This alone, though suggestive, does not imply any constraints, since there is no argument for these anomalous dimensions to have any particular sign. However, if this proportionality were found to hold also at large $n$, then causality would imply the full Hofman-Maldacena constraints $n_{s,f}\geq 0$. Significant progress has been made recently on this limit of the lightcone bootstrap for scalars \cite{Kaviraj:2015xsa,Kaviraj:2015cxa}.

 In fact, the actual calculation of anomalous dimensions may not even be necessary. It would be sufficient to argue that, in the shockwave limit  $|1-\bz|\ll|1-z|\ll1 $, the scalar and fermion pieces of the stress tensor block (or some combinations thereof) decompose into orthogonal families of operators in the dual channel. The lightcone bootstrap is one way to perform this decomposition explicitly, but perhaps there is a less explicit way to make this argument.

Similar comments apply to $\langle TT\psi\psi\rangle$. In that case there are three families of composite operators: symmetric $\tableau{3}\cdots$, mixed type $\tableau{3 1}^{\dots}$, and mixed type $\tableau{4 2}^{\dots}$.

\bigskip

\bigskip

\textbf{Acknowledgments}
We thank J.~Kaplan, Z.~Komargodski, E.~Perlmutter, D.~Poland, M.~Walters, and Z.~Zhiboedov for many useful discussions of spinning correlators, and are grateful to the authors of \cite{Li:2015itl} for sharing their results prior to publication. The work of TH and SJ is supported by DE-SC0014123. The work of SK is supported by the NSF grant PHY-1316222.

\appendix

\section{Review of spinning correlators}\label{s:structure}

In this section we review the structure of 2-, 3-, and 4-point functions of spinning operators in CFT, using the techniques of \cite{cppr1,cppr2} which we follow closely. Throughout the paper, $\psi$ and $\phi$ are scalars, $T$ is the stress tensor, and $J$ is a spin-1 conserved current. $O$ is any symmetric traceless tensor operator.

\subsection{Embedding space notation}
In index-free notation, a symmetric tensor operator at a point $x_i^\mu \in R^d$ is represented as
\be
O_{\mu\nu\cdots}(x_i) \to O(P_i, Z_i)
\ee
where $P_i, Z_i \in R^{d,2}$. $P$ encodes the position and $Z$ encodes the polarization tensor. To translate to physical space, one simply sets
\be
P_i  = (P^+_i, P^-_i, P^\mu_i) \to (1, x_i^2, x_i^\mu) , \qquad Z_i \to (0, 2 x_i \cdot \epsilon_i, \epsilon_i^\mu)
\ee
with inner product $-dX^+ dX^- + \delta_{\mu\nu}dX^\mu dX^\nu$. We choose null polarizations, $\epsilon^2=0$, unless specified otherwise; then $O(P_i, Z_i)$ represents 
\be\label{epsj}
\epsilon_i^\mu \epsilon_i^\nu \cdots O_{\mu\nu\cdots}(x_i) \ .
\ee
For traceless symmetric tensors, the null contractions \eqref{epsj} completely determine the full operator $O_{\mu\nu\cdots}$, so it is possible to translate between null polarizations $\epsilon^\mu\epsilon^\nu\cdots$ and arbitrary polarizations $\varepsilon^{\mu\nu\cdots}$. The components of the full tensor are 
\be\label{tensorizer}
O_{\mu_1\dots\mu_\ell} = \frac{1}{\ell! (d/2-1)_{\ell}}D_{\mu_1}\cdots D_{\mu_\ell} \left[ \epsilon^{\nu_1}\cdots \epsilon^{\nu_\ell} O_{\nu_1 \cdots \nu_\ell}\right]
\ee
where
\be\label{translator}
D_\mu = \left(\frac{d}{2}-1 + \epsilon \cdot \frac{\p}{\p \epsilon}\right) \frac{\p}{\p \epsilon^\mu} - \frac{1}{2}\epsilon_\mu \frac{\p^2}{\p \epsilon \cdot \p\epsilon} \ .
\ee
In practice, for spin 1 one simply replaces $\epsilon^\mu \to \varepsilon^\mu$, \ie the null result applies also to non-null polarization vectors, while for spin 2, we replace $\epsilon^\mu \epsilon^\nu \to \varepsilon^{\mu\nu}$ with $\varepsilon^{\mu}_{\ \mu} = 0$.

\subsection{Building blocks}
The building blocks of conformal correlators are the combinations
\bea\label{hvdef}
H_{ij} &=& -2\left[(Z_i \cdot Z_j)(P_i \cdot P_j) - (Z_i \cdot P_j)(Z_j \cdot P_i)\right] = -2 x_{ij}\sdot \e_j x_{ij}\sdot\e_i + x_{ij}^2 \epsilon_i \sdot \epsilon_j\notag\\
V_{i,jk} &=& \frac{(Z_i \cdot P_j)(P_i \cdot P_k) - (Z_i \cdot P_k)(P_i \cdot P_j)}{P_j \cdot P_k} = \frac{1}{x_{jk}^2}\left(x_{ij}^2 x_{ik}\sdot \epsilon_i - x_{ik}^2 x_{ij}\sdot \epsilon_i\right)
\eea
with $x_{ij} = x_i - x_j$. To shorten the notation we also use
\begin{align}
V_1 &= V_{1,23}  & V_1' &= V_{1,24} &
V_2 &= V_{2,31}  & V_2' &= V_{2,14}\\
V_3 &= V_{3,12}  & V_{3}' &= V_{3,42} & 
V_4 &= V_{4,12} & V_4' &=V_{4,13} \ .\notag
\end{align}
Define 
\be
P_{ij} = -2P_i \cdot P_j =x_{ij}^2 , \qquad K_{ijk} = \frac{P_{jk}}{P_{ik}} \ ,
\ee
and the standard cross-ratios are
\be
u = \frac{P_{12}P_{34}}{P_{13}P_{24}} \ , \quad v=\frac{P_{14}P_{23}}{P_{13}P_{24}} \ ,
\ee
which will often be traded for $z,\bz$ defined by
\be\label{uvzz}
u = z\bz , \qquad v=(1-z)(1-\bz) \ .
\ee
The building blocks are not all independent. In any spacetime dimension, they obey
\be
V_{3,24} + u V_{3,12}- v V_{3,14} = u V_{2,34} - v V_{2,14} + V_{2,13} = u V_{1,34} + v V_{1,23} - V_{1,24} = 0 \ ,
\ee
and there are further relations in low dimensions.

\subsection{2-point functions}

The two-point function of a primary of dimension $\Delta$ and spin $\ell$,
\be
G(P_1, Z_1; P_2, Z_2)  = \langle O(P_1, Z_1) O(P_2, Z_2) \rangle
\ee
is fixed by conformal invariance up to an overall normalization.  The only building block available for two $P$'s is $H_{12}$ in \eqref{hvdef}, and the answer must contain $\ell$ copies each of $\e_1$ and $\e_2$. Inserting a factor of $P_{12}$ to get the correct scaling weight, this determines
\be\label{deftwo}
G(P_1, Z_1; P_2, Z_2) = c_O \frac{(H_{12})^\ell}{(P_{12})^{\Delta+\ell}} \ .
\ee
Here $c_O$ is a positive normalization constant. The sign is fixed by requiring
\be
(\epsilon^\mu \epsilon^\nu \cdots)(\e^{*\alpha} \e^{*\beta} \cdots ) \langle O_{\mu\nu\dots}(-x) O_{\alpha\beta\dots}(x)\rangle  > 0
\ee
where $\e\cdot x=0$. This is a special case of reflection positivity.

\newcommand{\tcurly}[3]{\left\{ #1 \quad #2 \quad #3 \right\} }
\newcommand{\squareb}[9]{{\tiny \left[\begin{array}{ccc}
#1 & #2 & #3\\
#4 & #5 & #6\\
#7 & #8 & #9
\end{array}\right]}}
Throughout the paper, we set $c_O=1$ for scalar operators, but use the canonical normalization of currents and stress tensors.

\subsection{3-point functions}
A full 3-point function is a sum over structures built from $H_{ij}$ and $V_{i,jk}$ that can appear with independent coefficients.   
Each combination of $H$'s and $V$'s with the appropriate number of $\epsilon_i$'s, together with factors of $P_{ij}$ to get the correct scaling weights, is an independent structure.  Thus a basis of structures is 
\be
 \frac {V_{1,23}^{m_1}V_{2,31}^{m_2}V_{3,12}^{m_3}H_{12}^{n_{12}}H_{13}^{n_{13}}H_{23}^{n_{23}}}{ (P_{12})^{\half(h_1+h_2-h_3)}(P_{13})^{\half(h_1+h_3-h_2)}(P_{23})^{\half(h_2+h_3-h_1)}} \ ,
\ee
where the $n_{ij}$ are nonnegative integers satisfying
\be\label{msat}
m_1 \equiv \ell_1 - n_{12} -n_{13}\geq 0 , \quad m_2 \equiv \ell_2 - n_{12}-n_{23}\geq 0 , \quad m_3 \equiv \ell_3 - n_{13}-n_{23} \geq 0  \ ,
\ee
and we have defined
\be
h_i = \Delta_i+\ell_i \ .
\ee

\subsection{4-point functions}
A 4-point function of operators with spin can be written
\be\label{genfour}
\left( P_{24}\over P_{14}\right)^{\half(h_1-h_2)}\left(P_{14}\over P_{13}\right)^{\half(h_3-h_4)}(P_{12})^{-\half(h_1+h_2)}(P_{34})^{-\half(h_3+h_4)} \sum_A f_A(z,\bz) Q^{(A)}(H_{ij}, V_{i,jk}) \ ,
\ee
where recall $h_i \equiv \Delta_i + \ell_i$.
Each structure $Q^{(A)}$ is a polynomial in the building blocks \eqref{hvdef} in which each $Z_i$ appears $\ell_i$ times. The coefficients are functions of the cross-ratios $z$ and $\bz$ (or equivalently $u$ and $v$), not fixed by conformal symmetry.

\subsection{Conservation}

The divergence operator in embedding space is 
\be
\del_{P,Z} = \frac{\p}{\p P_M}\left[\left(\frac{d}{2}-1 + Z \cdot \frac{\p}{\p Z}\right) \frac{\p }{\p Z^M} - \frac{1}{2}Z_M \frac{\p^2}{\p Z\cdot\p Z}\right] \ .
\ee
Up to contact terms, a correlator with a conserved operator at the point $(P,Z)$ must vanish upon applying this operator.

\section{3-point functions of conserved currents}\label{s:currents}
In this appendix we collect our notation and conventions for the couplings in $\langle JJT\rangle$ and $\langle TTT\rangle$. 

Correlators of conserved currents in $d$-dimensional CFT were first found in \cite{Osborn:1993cr} (see also \cite{Erdmenger:1996yc}). We use the embedding space formalism \cite{cppr1,cppr2}. Useful formulae for comparison, in similar notation, can be found in \cite{Zhiboedov:2012bm,Zhiboedov:2013opa,Li:2015itl}.

\subsection{$JJT$}
Applying the technology above to $\langle JJT\rangle$ gives
\begin{align}\label{jjtbasis}
\langle J(P_1,Z_1)J(P_2,Z_2)&T(P_3,Z_3)\rangle \\
& = \frac{ \alpha_1 V_{1} V_{2} V_{3}^2 + \alpha_2 H_{12} V_{3}^2 +  \alpha_3 ( H_{23} V_{1}V_{3}+H_{13}V_{2}V_{3} ) + \alpha_5 H_{13}H_{23}}{P_{12}^{\frac{d}{2}-1}P_{13}^{\frac{d}{2}-1}P_{23}^{\frac{d}{2}+1}} \ .\notag
\end{align}
This can also be written in the free field basis 
\be\label{freej}
\langle JJT \rangle = n_s \langle JJT\rangle_{scalar} + n_f \langle JJT \rangle_{fermion}
\ee
where the coupling constants are
\begin{equation}
\begin{split}
&\alpha_1 = n_s \frac{d-2}{2(d-1)}-8 n_f,~~ \alpha_2= -4 n_f -  \frac{n_s}{2(d-1)} \\
&\alpha_3 = -4n_f - \frac{n_s}{d-1},~~\alpha_5= \frac{n_s}{(d-1)(d-2)}.
\end{split}\end{equation}
This defines our normalization for the structures in \eqref{freej}. In a theory with (integer) $N_s$ free real scalars and $N_f$ free Dirac fermions, all of charge 1, 
\be
n_s = \frac{d}{S_d^3}N_s , \quad n_f = \frac{d }{S_d^3}\frac{2^{\left \lfloor{d/2}\right \rfloor }}{4}N_f \ ,
\ee
where $S_d = \frac{ 2\pi^{d/2} }{ \Gamma(d/2)}$. The Ward identity relates one combination of the 3-point couplings to the 2-point function defined in \eqref{deftwo}:
\be
c_J  = \frac{S_d}{d}\left(4n_f + \frac{n_s}{d-2}\right) \ .
\ee

\subsection{$TTT$}
Conformal invariance and permutations fix
\be
 \langle T(P_1,Z_1)T(P_2,Z_2) T(P_3,Z_3)\rangle= \frac{\sum_{i=1}^{5}\alpha_i S_i}{ P_{12}^{1+\frac{d}{2} } P_{13}^{1+\frac{d}{2} } P_{23}^{1+\frac{d}{2}} }
\ee
where
\begin{align}
&S_1=  V_1^2 V_2^2 V_3^2 ,~~ S_2= V_1 V_2 V_3 \left( H_{23} V_1+H_{13} V_2+H_{12} V_3\right) \\
& S_3=\left(H_{12} H_{23} V_1 V_3+ H_{13} V_2 \left( H_{23} V_1+ H_{12} V_3  \right)  \right),~~ S_4= H_{12} H_{13} H_{23}\notag\\
&S_5=H_{23}^2 V_1^2+ H_{13}^2 V_2^2 + H_{12}^2 V_3^2.\notag
\end{align}
The translation to the free-field basis
\be\label{freet}
\langle TTT \rangle = \tn_s \langle TTT\rangle_{scalar} + \tn_f \langle TTT \rangle_{fermion}+ \tn_v\langle TTT \rangle_{vector}
\ee
is
\begin{align}
\alpha_1 &= 128 d^2 \tn_f - \frac{8d^2(d-2)^3}{(d-1)^3}\tn_s - 8192\tn_v \\
\alpha_2 &= 64d(d-2)\tn_f + \frac{32(d-2)^2d^2}{(d-1)^3}\tn_s-8192\tn_v\notag\\
\alpha_3 &= -128d\tn_f - \frac{64d^2(d-2)}{(d-1)^3}\tn_s-4096\tn_v\notag\\
\alpha_4 &=\frac{64d^2}{(d-1)^3}\tn_s - \frac{4096}{d-2}\tn_v\notag\\
\alpha_5 &=-64d\tn_f - \frac{16d(d-2)^2}{(d-1)^3}\tn_s-2048\tn_v\notag
\end{align}
In a theory of (integer) $N_s$ free real scalars, $N_f$ free Dirac fermions, or $N_v$ free vectors,
\be
\tn_s = \frac{d}{S_d^3}N_s , \quad \tn_f = \frac{d }{S_d^3}\frac{2^{\left \lfloor{d/2}\right \rfloor }}{4}N_f \ , \quad \tn_v = \frac{d^3(d-2)\Gamma(d-1)}{256\pi^d(d-3)S_d} N_v \ .
\ee
The coefficient of the 2-point function, related by the Ward identity, is
\be
c_T = 128S_d\left(\tn_f + \frac{1}{2(d-1)}\tn_s + \frac{16(d-3)}{d(d-2)}\tn_v\right) \ .
\ee
In $d=4$, the ratio of anomaly coefficients is
\be
\frac{a}{c} = \frac{\tn_s + 11 \tn_f + 62 \tn_v}{3\tn_s + 18 \tn_f + 36 \tn_v} \ .
\ee

\section{Conformal blocks for stress tensor exchange}

Conformal blocks are computed following \cite{cppr2}. This method produces spinning conformal blocks as differential operators acting on the conformal blocks with external scalars, $g_{\Delta,\ell}^{\Delta_{12},\Delta_{34}}(u,v)$.  Many details of the intermediate steps for $JJ \to T \to \psi\psi$ and $TT \to T \to \psi\psi$ appear in \cite{Li:2015itl}. We will quote the results only for $JJ\psi\psi$, as the expressions for $TT\psi\psi$ are too long to be written usefully.

Consider the correlator
\be
\langle J(P_1,Z_1)J(P_2,Z_2)\psi(P_3)\psi(P_4)\rangle \ ,
\ee
in $d=4$.  For the exchange of a general symmetric traceless tensor $JJ \to O \to \psi\psi$, the full partial wave (\ie including all prefactors as well as the conformal blocks) has the form
\be\label{fg0}
W_O  = \frac{c_{\psi \psi O}}{c_O(P_{34})^{\Delta_\psi}(P_{12})^4} \left( f_1^O H_{12} + f_2^O V_1V_2 + f_3^O V_1'V_2 + f_4^O V_1V_2' + f_5^OV_1' V_2'\right) \ .\ee
The coefficient functions are
\bea\label{fj}
f_1^{O} &=& \big[-\tfrac{1}{3}(n_s+12n_f) + \tfrac{1}{6}n_s u \p_u] g\\
f_2^O &=& -\tfrac{1}{3}n_s  v(\p_v + v \p_v^2 + u  \p_u \p_v)g \notag\\
f_3^O &=& \tfrac{1}{3}n_s(\p_v + v \p_v^2)g +\tfrac{4}{3} n_f\big[ (1-v)(\p_v+v\p_v) - u\p_u - 2 uv \p_u \p_v - u^2 \p_u^2\big]g\notag\\
f_4^O &=& -\tfrac{1}{3}n_s v (v\p_v + v^2 \p_v^2 + u \p_u + 2 u v \p_u \p_v + u^2 \p_u^2)g \notag\\
& & \qquad + \tfrac{4}{3}n_f v\big[ (1-v)(\p_v+v\p_v) - u \p_u-2uv\p_u\p_v - u^2\p_u^2\big]g\notag\\
f_5^O &=& -f_2^O\notag \ ,
\eea
where $g$ is the scalar block, $g \equiv g_{\Delta_O, \ell_O}^{0,0}(u,v)$. This expression passes a number of checks:  It is conserved, has the appropriate permutation symmetries, and in the Euclidean limit $u,v \to 0$, it has the correct leading behavior 
\be
W_O \approx \langle J(x_1)J(x_2)| T_{\mu\nu}\rangle \langle T_{\sigma\rho} | \psi(x_3)\psi(x_4)\rangle K^{\mu\nu,\sigma\rho}
\ee 
where $K = \langle TT\rangle^{-1}$. This last check is a completely independent calculation of the leading term, using only the formulas in \cite{Osborn:1993cr}, so serves as a nontrivial check that \cite{cppr2} has been implemented correctly.

To find the lightcone block for stress tensor exchange, we trade $(u,v)$ for $(z,\bz)$ using \eqref{uvzz} and plug in the scalar lightcone block for stress tensor exchange in $d=4$,
\be
g \approx \frac{\bz}{4} z^3 \, _2F_1(3,3,6,z) \ .
\ee
In the shockwave calculation, as explained in \cite{paper1}, this hypergeometric function must be evaluated after $1-z \to ( 1- z)e^{-2\pi i}$. Explicitly, for the results in section \ref{ss:jjres} we use \eqref{fg0}, \eqref{fj} with
\be
g = \frac{15\bz}{2z^2}\left(3z(z-2) -(6-6z+z^2)\left[\log(1-z)-2\pi i\right]\right) \ .
\ee
(Note that here we are using $s$-channel notation but section \ref{ss:jjres} is in $t$-channel notation, so it is also necessary to swap $u \leftrightarrow v$ and relabel the tensor structures.)

The calculation for $TT \to O \to \psi\psi$ is similar. The only slight conceptual difference is that for spin $>1$, the embedding space approach directly gives the answer only for null polarizations, $\varepsilon_{\mu\nu} = v_\mu v_\nu$ with $v^2=0$. The causality constraints require general polarizations. The results quoted in section \ref{s:shock} were obtained from the embedding-space result by applying the procedure described around \eqref{tensorizer} to constructed the full tensor.

\end{spacing}

\end{document}